\documentclass[twocolumn,prb]{revtex4}
\usepackage{graphicx,amsmath,amssymb}

\begin{document}

\title{Lissajous curves and semiclassical theory: The two-dimensional
harmonic oscillator}
\author{Roland Doll}
\email{roland.doll@physik.uni-augsburg.de}
\author{Gert-Ludwig Ingold}
\affiliation{Institut f{\"u}r Physik, Universit{\"a}t Augsburg,
Universit{\"a}tsstra{\ss}e 1, D-86135 Augsburg, Germany}

\begin{abstract}
The semiclassical treatment of the two-dimensional harmonic oscillator
provides an instructive example of the relation between classical motion and 
the quantum mechanical energy spectrum. We extend previous work on the
anisotropic oscillator with incommensurate frequencies and the isotropic
oscillator to the case with commensurate frequencies for which the Lissajous
curves appear as classical periodic orbits. Because of the three different
scenarios depending on the ratio of its frequencies, the two-dimensional
harmonic oscillator offers a unique way to explicitly analyze the role of
symmetries in classical and quantum mechanics.
\end{abstract}
\maketitle

\section{Introduction}
In 1940 J.\ M.\ Jauch and E.\ L.\ Hill discussed the relation between
symmetries and degeneracies in the spectra of quantum systems. As examples
they considered the hydrogen problem, the two-dimensional Kepler problem, and 
the isotropic harmonic oscillator, but the two-dimensional anisotropic
harmonic oscillator resisted their analysis. The authors concluded ``We feel,
however, that the difficulties encountered in this relatively simple
problem throw question on the true interpretation of classical
multiply-periodic motions in quantum mechanics.'' \cite{jauch40} 

Although the problem is now understood on a group theoretical level,
\cite{louck73,krame75,moshi75} the 
remark by Jauch and Hill motivates us to study the relation
between the motion of a classical anisotropic oscillator with commensurate
frequencies and its quantum mechanical energy spectrum. This link is
provided by semiclassical theory \cite{gutzw98} which relates the density 
of states to properties of the classical periodic orbits. We will restrict
the discussion to the two-dimensional case where harmonic potentials 
already provide three different scenarios for periodic orbits.

\begin{figure}[t]
\includegraphics[width=\columnwidth]{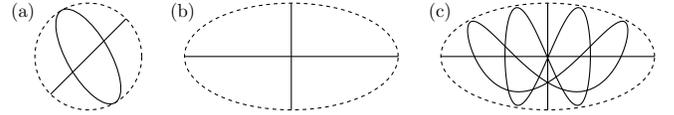}
\caption{\label{fig:traj}Examples of the periodic orbits of a two-dimensional
harmonic oscillator. The dashed line corresponds to an equipotential line.
(a) Isotropic harmonic oscillator. (b) Anisotropic harmonic oscillator with
incommensurate frequencies. In this case only the two orbits shown are
periodic. (c) Anisotropic harmonic oscillator with commensurate
frequency ratio 1:2. In contrast to the case with incommensurate
frequencies, Lissajous curves become possible.}
\end{figure}

For the isotropic harmonic oscillator all
orbits are periodic and constitute a continuous family of which two
representatives are depicted in Fig.~\ref{fig:traj}(a). Any two distinct
periodic orbits can be deformed continuously into each other via other
periodic orbits. The situation is different for an anisotropic
harmonic oscillator with incommensurate frequencies where only the two
normal modes shown in Fig.~\ref{fig:traj}(b) correspond to periodic orbits.
The most interesting classical dynamics is exhibited by the harmonic
oscillator with commensurate frequencies of which an example is shown in
Fig.~\ref{fig:traj}(c). Here, Lissajous curves appear as additional
periodic orbits that form a continuous family. These trajectories, which
will play an important role in the following, take their name after Jules
Antoine Lissajous who gave an extensive discussion of these curves in a
study aimed at a comparison of the relative frequencies
of two tuning forks. \cite{lissa57} Previous work addressing these curves
should not be forgotten.\cite{crowe81}

It has been proven that semiclassical theory gives the exact
expression for the density of states of an isotropic\cite{creag96} as well
as for the incommensurate harmonic oscillator.\cite{brack95} 
An interesting property of the
harmonic oscillator is that the appropriate limit from incommensurate to
commensurate frequencies yields the correct density of states even for the
commensurate harmonic oscillator,\cite{brack95} although only two periodic 
orbits are explicitly taken into account. Although taking such a limit
considerably simplifies the calculation, it is unsatisfying
that a full analysis of the density of states based on all the classical
periodic orbits is lacking. Furthermore, a complete account for the 
symmetries of the commensurate anisotropic
harmonic oscillator is not possible without the Lissajous curves. As we
will show, the proper treatment of all the periodic orbits of the commensurate 
anisotropic harmonic oscillator yields the exact density of states and
the exact symmetry reduced densities of states. Our analysis thus closes
the last lacuna left in the semiclassical treatment of the two-dimensional
harmonic oscillator. This system now lends itself well to illustrating the 
relation between classical dynamics and the symmetry properties of the
quantum mechanical energy spectrum. 
\section{The energy spectrum in terms of classical
mechanics: the incommensurate harmonic oscillator}
\label{sec:i-AHO} Among the three types of two-dimensional harmonic
oscillators depicted in Fig.~\ref{fig:traj}, the relation between the
quantum mechanical density of states and classical properties is most
readily established for the incommensurate harmonic oscillator.
The Hamiltonian of a particle of mass $m$ moving
in a two-dimensional harmonic potential with eigenfrequencies $\omega_1$ and
$\omega_2$ is given by
\begin{equation}
H = \frac{p_1^2}{2m} + \frac{m}{2}\omega_1^2q_1^2 
+ \frac{p_2^2}{2m} + \frac{m}{2}\omega_2^2q_2^2.
\label{eqnHamiltonian}
\end{equation}
We first use the fact that the total Hamiltonian
$H$ can be decomposed into two contributions $H_1(q_1,p_1)$ 
and $H_2(q_2,p_2)$ describing two uncoupled one-dimensional 
harmonic oscillators.
From the classical density of states of the $i$th normal mode,
\begin{equation}
\bar\rho_i(E)=\!\int\frac{dq_idp_i}{2\pi\hbar}
\delta \Big(E-\frac{p_i^2}{2m}-\frac{m}{2}\omega_i^2q_i^2\Big)
= \frac{1}{\hbar\omega_i},
\end{equation}
the classical density of states of the
two-dimensional oscillator is obtained by the convolution
\begin{equation}
\label{eqnClassicalDOS}
\bar\rho(E)=\!\int_0^E dE'\bar\rho_1(E-E')\bar\rho_2(E')=
\frac{E}{\hbar^2\omega_1\omega_2}.
\end{equation}

The quantum mechanical energy spectrum is
given by
\begin{equation}
E_{\nu_1,\nu_2} = \hbar\omega_1\Big(\nu_1 + \frac{1}{2}\Big) 
+ \hbar\omega_2\Big(\nu_2 + \frac{1}{2}\Big),
\label{eqnQmEnergies}
\end{equation}
where $\nu_1, \nu_2 = 0, 1, 2,\dots$ The corresponding quantum mechanical 
density of states is
\begin{equation}
\label{eq:exactdos}
\rho(E) = \sum_{\nu_1,\nu_2=0}^\infty 
\delta(E - E_{\nu_1,\nu_2}).
\end{equation}

Equation~\eqref{eq:exactdos} can be interpreted 
in terms of the classical properties of the incommensurate 
harmonic oscillator if it is appropriately re-expressed. We perform a Laplace
transform of Eq.~(\ref{eq:exactdos})
and find the partition function of the incommensurate 
harmonic oscillator
\begin{equation}
\label{eqnPartitionFunction-AHO}
\begin{aligned}
\mathcal{Z}_\beta &=\!\int_0^{\infty}\!dE\,\rho(E)\exp(-\beta E)\\
&= \frac{1}{4\sinh\Big(\displaystyle\frac{\hbar\beta\omega_1}{2}\Big)
\sinh\Big(\displaystyle\frac{\hbar\beta\omega_2}{2}\Big)}.
\end{aligned}
\end{equation}
The inversion of the Laplace transform by means of a contour 
integration yields three kinds of contributions from the various
poles on the imaginary axis. The double pole at $\beta=0$ gives rise to the
classical density of states (\ref{eqnClassicalDOS}), and the poles at
$\beta=\pm i2\pi n/\hbar\omega_1$ and $\beta=\pm i2\pi n/\hbar\omega_2$ with
$n=1,2,\dots$ lead to contributions responsible for the difference between 
the classical and quantum density of states. The latter can thus be expressed
as
\begin{equation} \label{eqnClassicalDOSPlusFluctuations}
\rho(E) = \bar\rho(E) + \delta \rho_1(E) + \delta \rho_2(E).
\end{equation}
After evaluating the residues, we obtain
\begin{equation}
\label{eqnScDosi-AHO}
\delta \rho_1(E) = \frac{1}{\pi\hbar}
\sum_{r=1}^{\infty} T_1\frac{\cos\Big(\displaystyle
r\frac{2\pi E}{\hbar\omega_1} -\frac{\pi}{2}\sigma_{r,1}\Big)}
{2|\sin(\pi r\omega_2/\omega_1)|},
\end{equation}
with
\begin{equation}\label{eqnMaslovIndexi-AHO}
\sigma_{r,1} = 2r+1+2\Big[r\frac{\omega_2}{\omega_1}\Big].
\end{equation}
The quantities $\delta \rho_2(E)$ and $\sigma_{r,2}$ are expressed in the
same way with the indices
$1$ and $2$ interchanged. We have written the expressions in a form that
anticipates the following discussion. In particular, we have introduced the
periods
$T_i=2\pi/\omega_i$ of the two normal modes and phases $\sigma_{r,i}$. 
In Eq.~(\ref{eqnMaslovIndexi-AHO}) $[x]$ denotes the integer part of $x$.

We expect that it should be possible to relate the terms in 
$\delta \rho_i$ to the properties of the two periodic orbits shown in 
Fig.~\ref{fig:traj}(b). As a motivation we remark that within the Feynman
path integral approach\cite{feynm48} all paths joining two points in a
given time contribute to the corresponding quantum propagator a phase
factor $\exp(iS/\hbar)$, where $S$ is Hamilton's principal function of 
the path. For $S\gg\hbar$, we can apply the semiclassical 
approximation\cite{gutzw98,remark} and find that the path integral is
dominated by the stationary points of $S$, that is, the classical paths.
Within this approximation even the contribution of quantum fluctuations
can be expressed in terms of properties of the classical 
motion.\cite{brack97,ingol02}

The propagator can be related to the density of states by means of the Green 
function $G(E)=(E-H)^{-1}$, the Fourier transform of the
propagator. From the relation 
\begin{equation}
\label{eqnDiracIdentity}
\begin{aligned}
\rho(E) &= -\frac{1}{\pi} \lim_{\epsilon \to 0^+} \text{Im}\,
(\text{Tr}\,[G(E+i\epsilon)])\\
&= -\frac{1}{\pi} \lim_{\epsilon \to 0^+} \text{Im}
\sum_{\nu_1,\nu_2}\frac{1}{E+i\epsilon-E_{\nu_1,\nu_2}},
\end{aligned}
\end{equation}
we see that $\rho(E)$ involves the trace over the Green function, so that
the initial and final points of the contributing classical paths 
coincide. Equation~(\ref{eqnDiracIdentity}) contains contributions
from very short orbits which yield $\bar\rho(E)$ and periodic orbits of
finite length contribute to
$\delta \rho(E)=\delta \rho_1(E)+\delta \rho_2(E)$. The latter relation
is given by the Gutzwiller trace formula 
\cite{gutzw90,brack97} of which Eq.~(\ref{eqnScDosi-AHO}) is an
example.

These considerations indicate that within the semiclassical approximation 
the density of states can be expressed in terms of classical properties, 
but such an expression cannot be expected in general.
However, it was shown by Brack and Jain\cite{brack95} that the
incommensurate harmonic oscillator is one of the rare cases
where the Gutzwiller formula yields the exact density of states. As a
consequence, it should be possible to interpret the various terms in 
Eq.~(\ref{eqnScDosi-AHO}) by means of a classical periodic orbit. We
will check now that this is the case.

We have already mentioned that the prefactor $T_1$ in
Eq.~\eqref{eqnScDosi-AHO} represents the period of the first normal mode.
The argument of the cosine function in the numerator contains the classical
action 
\begin{equation}
W_1(E) = \!\oint dq_1 p_1 = \frac{2\pi E}{\omega_1}
\label{eqnAction}
\end{equation}
of the fundamental periodic orbit along the $q_1$ axis for a given energy $H_1 =
E$; we sum over all repetitions counted by $r$.

The denominator is determined by the lateral stability of the periodic
orbit as the following argument shows. The equations of motion of a
harmonic oscillator can easily be solved, and we find
\begin{equation}
\begin{pmatrix}
q_i(t) \\
p_i(t) \\
\end{pmatrix}
= \mathsf{X}_i(t)
\begin{pmatrix}
q_i(0) \\
p_i(0)
\end{pmatrix}
\end{equation}
with the matrizant
\begin{equation}
\label{eqnMatrizant}
\mathsf{X}_i(t)
= \begin{pmatrix}
\cos(\omega_i t) & \displaystyle\frac{\sin(\omega_i t)}{m\omega_i} \\
-m\omega_i\sin(\omega_i t) & \cos(\omega_i t) \\
\end{pmatrix}.
\end{equation}

We again consider the periodic orbit along the $q_1$ axis 
which is specified by the initial conditions $q_2(0)=0$ and $p_2(0)=0$. We
are interested in the evolution of a small lateral perturbation during 
$r$ oscillations. After the time $r T_1$, we find a deviation 
\begin{equation}
\begin{pmatrix}
\delta q_2 \\
\delta p_2
\end{pmatrix}
= [
\mathsf{X}_2(r T_1) - \openone]
\begin{pmatrix}
q_2(0)\\
p_2(0)
\end{pmatrix}
\end{equation}
in the direction orthogonal to the periodic orbit with respect to the perturbed 
initial condition. The matrix $\mathsf{X}_2(r T_1)$ thus measures the stability 
during $r$ repetitions of the first normal mode against a small initial 
perturbation. It enters the 
denominator of Eq.~(\ref{eqnScDosi-AHO}) by means of its property
\begin{equation} \label{eqnDenominator}
\sqrt{|\det[\mathsf{X}_2(rT_1) -\openone]|} = 2 | \sin(\pi r\omega_2/\omega_1)|.
\end{equation}

The last quantity requiring interpretation is the Maslov index
$\sigma_{r,1}$ defined by Eq.~(\ref{eqnMaslovIndexi-AHO}). It collects all
phases appearing in the evaluation of the stationary phase integral of the
first periodic orbit. Because the derivation of
Eq.~(\ref{eqnMaslovIndexi-AHO}) within a semiclassical approach is more
demanding, we refer the reader interested in the details to the
literature.\cite{brack95}

We started from an exact expression for the density of states of the
incommensurate harmonic oscillator and arrived at a
semiclassical interpretation. In contrast, in the following two sections,
we will go the opposite way and first derive a semiclassical expression for
the density of states of the isotropic harmonic oscillator and the
commensurate anisotropic harmonic oscillator by exploiting the system's classical
properties. At the end of each section, we will then compare with the exact
quantum mechanical result.

\section{Semiclassical theory in the presence of a continuous symmetry: the
isotropic harmonic oscillator}
\label{sectIHO}
As we have seen, the semiclassical density of states is intimately related 
to the classical periodic orbits. Because the incommensurate 
harmonic oscillator exhibits only two such orbits, its semiclassical density
of states is relatively simple to determine.

As the next step toward a semiclassical understanding of the commensurate
anisotropic harmonic oscillator, we consider the more complex case of equal frequencies,
$\omega_1=\omega_2=\omega$. The fundamental periods of oscillations along
the $q_1$- and $q_2$-axes are equal, $T_1 = T_2 = T = 2 \pi/\omega$ and
from Eq.~(\ref{eqnMatrizant}) we find $\mathsf{X}_2(rT_1) =
\mathsf{X}_1(rT_2) =
\openone$ for any integer $r$, which is related to the fact that any starting
point on an energy surface $H=E$ in the four-dimensional phase space leads
to a periodic orbit. In particular, the oscillations along the $q_1$- and
$q_2$-axes are still periodic and should contribute to the semiclassical
density of states as was the case in Sec.~II. However, the denominators in 
Eq.~(\ref{eqnScDosi-AHO}) reflecting the stability of the orbits vanish.
The resulting divergences indicate that the 
evaluation of the propagator within the stationary phase approximation fails 
because the stationary points are no longer isolated but form a continuous
family. A small perturbation of the initial conditions thus leads to another 
periodic orbit close to the original one. It is even possible to continuously 
deform the two Lissajous curves depicted in Fig.~\ref{fig:traj}(a) into
each other. The existence of a family of periodic orbits is a consequence of 
the presence of a continuous symmetry that maps a given orbit to a neighboring 
one. Hence, the relation between the classical properties and the density of 
states discussed in the Sec.~II has to be modified. 

To understand the necessary changes, we consider the phase
space symmetries of the isotropic harmonic oscillator and define
the quantities
\cite{golds80,mcint59}
\begin{subequations} \label{eqnConservedQuantities}
\begin{align}
J_1 &= \frac{1}{2\omega}\Big(\frac{p_1p_2}{m}+m\omega^2q_1q_2\Big)\\
J_2 &= \frac{1}{2}(q_1 p_2- q_2 p_1)\\
J_3 &= \frac{1}{2\omega}\Big(\frac{p_1^2}{2m} + \frac{m}{2}\omega^2 q_1^2 
-\frac{p_2^2}{2m}-\frac{m}{2}\omega^2 q_2^2\Big),
\label{eq:j3}
\end{align}
\end{subequations}
for which we find the Poisson brackets
\begin{equation}
\label{eqnPoisson}
\{J_i,J_j\}=\epsilon_{ijk}J_k,
\end{equation}
where $\epsilon_{ijk}$ is the completely antisymmetric tensor. Two of the 
$J_i$ have an obvious physical meaning. $J_2$ is proportional to the 
angular momentum and generates rotations around the axis perpendicular to the 
$q_1$-$q_2$ plane. It thus accounts for the rotational symmetry of the 
isotropic oscillator in real space. $J_3$ represents the difference of the 
energies of the normal modes. $J_2$ and $J_3$ are conserved quantities as
can be verified explicitly by evaluating the Poisson bracket with the 
Hamiltonian of the isotropic harmonic oscillator. Because $J_1$ results from
the Poisson bracket of two conserved quantities, it is a constant of the
motion as well. We also have
\begin{equation}
\label{eqnCasimir}
J_1^2+J_2^2+J_3^2 = \Big(\frac{E}{2\omega}\Big)^2.
\end{equation}
From the quantum version of Eq.~(\ref{eqnPoisson}) 
\begin{equation}
[J_i,J_j]=i\hbar \epsilon_{ijk} J_k
\end{equation}
the relation to spin 1/2 is evident. Both the spin 1/2 and the
isotropic two-dimensional harmonic oscillator exhibit a SU(2) symmetry,
which indicates how the periodic orbits of the isotropic harmonic
oscillator can be classified. 

It is useful to recall the concept of a Bloch sphere. If we introduce 
spherical coordinates, every state of a spin 1/2 can be expressed in terms
of the orthogonal basis states $\vert{\uparrow}\rangle$ and
$\vert{\downarrow}\rangle$ as
\begin{equation}
\vert\psi\rangle = \cos(\theta/2)\vert{\uparrow}\rangle +
\sin(\theta/2)e^{i\varphi}\vert{\downarrow}\rangle,
\end{equation}
where $\theta$ runs from $0$ to $\pi$ and $\varphi$ from $0$ to $2\pi$.
This representation provides a one-to-one mapping between the spin states and 
the points on a sphere. In particular, the cartesian coordinates $x_i$ of a 
point on the Bloch sphere are proportional to the spin expectation values 
$\langle S_i\rangle$. We also note that the spin operators serve as
generators of rotations on the Bloch sphere so that an arbitrary state can
be generated by starting from a given initial state, for example,
$\vert{\uparrow}\rangle$.

In close analogy, there is a one-to-one correspondence between a point 
$\hat r$ on a unit sphere and a classical periodic orbit of the isotropic
harmonic oscillator. The orbit is uniquely characterized by the values of
the conserved quantities in Eq.~(\ref{eqnConservedQuantities}) which
can be expressed by
$\vec J=(E/2\omega)\hat r$. According to Eq.~(\ref{eq:j3}), the latitude on
the Bloch sphere determines the distribution of the energy between the two 
oscillation modes along the coordinate axes. The north and south pole of the 
Bloch sphere correspond to extremal values of $J_3$, that is, to an
oscillation along one of the coordinate axes. The equator represents all 
periodic orbits where the energy is equally shared between the two modes.
The position on the equator determines the relative phase of the two
oscillatory motions.

In four-dimensional phase space these three conserved quantities leave
one degree of freedom undetermined, which corresponds to the motion along
the periodic orbit generated by the Hamiltonian. In analogy to 
spin 1/2, all periodic orbits can be generated from a given periodic orbit by 
means of the conserved quantities in Eq.~(\ref{eqnConservedQuantities}).

A generalized trace formula that takes into account continuous symmetries
was given in Refs.~\onlinecite{creag91,creag92,murat03}. If we adapt Eq.~(5.1)
in Ref.~\onlinecite{creag92} to the special case of an isotropic harmonic
oscillator where all periodic orbits are members of a single two-parametric 
family, we find that the deviation from the classical density of states of
the semiclassical (sc) density of states is
\begin{equation} 
\label{eqnCreaghsTraceFormula}
\delta \rho_{\rm sc}(E) = \frac{TV|\vec J|}{2(\pi\hbar)^2}
\sum_{r=1}^\infty \cos\Big(r\frac{W}{\hbar} -
(\sigma_{r}+1)\frac{\pi}{2}\Big).
\end{equation}
As before, Eq.~\eqref{eqnCreaghsTraceFormula} is valid only for
positive energies. It resembles Eq.~(\ref{eqnScDosi-AHO}) in several
respects, but there are differences due to the existence of a family of
periodic orbits. We first remark that all members of the family share the
same action $W$, period
$T$, and Maslov indices $\sigma_r$ for $r$ repetitions of the periodic orbit.
Therefore, apart from the factor $T=2\pi/\omega$, which arises from the
integral along a periodic orbit, an additional factor $V\vert\vec J\vert$
appears as a result of the integration over the family. The volume occupied
by the family is given by the surface of the Bloch sphere $V=4\pi$. The
factor $\vert\vec J\vert=E/2\omega$ reflects our choice for the basis of the
Lie algebra characterizing the symmetry group. \cite{creag92,creag96} A
comparison of Eqs.~(\ref{eqnScDosi-AHO}) and (\ref{eqnCreaghsTraceFormula})
reveals that the factor derived from stability considerations in 
Eq.~(\ref{eqnDenominator}) is missing for the isotropic harmonic oscillator. 
This difference should be expected because every perturbation of a periodic 
orbit will lead to another periodic orbit so that in contrast to the 
incommensurate harmonic oscillator, the question of lateral stability for the
isotropic harmonic oscillator does not make sense.

We note that the
change in the symmetry properties prevents us from deriving the Maslov index
from the expression Eq.~(\ref{eqnMaslovIndexi-AHO}) by simply setting
$\omega_1 =
\omega_2$. The result for the incommensurate harmonic
oscillator has to be modified by an additional symmetry-induced
contribution\cite{creag96,doll04}
$\Delta\sigma=-2$ so that the Maslov index of the isotropic harmonic
oscillator becomes
\begin{equation}
\label{eqnMaslovIndexIHO}
\sigma_{r} = 4r-1.
\end{equation}

Substituting the expressions for the period $T$, the surface of the Bloch
sphere $V$, and the factor $\vert\vec J\vert$ given below
Eq.~\eqref{eqnCreaghsTraceFormula} as well as the action $W=2\pi
E/\omega$ and the Maslov index (\ref{eqnMaslovIndexIHO}) in the generalized
trace formula (\ref{eqnCreaghsTraceFormula}), we arrive at the semiclassical 
correction to the classical density of states for the isotropic harmonic 
oscillator
\begin{equation}
\label{eqnScDOSIHO}
\begin{aligned}
\delta \rho_{\rm sc} (E)
&= \frac{2 E}{\hbar^2\omega^2} \sum_{r=1}^\infty \cos \Big(
r \frac{2\pi E}{\hbar\omega} \Big) \\
&= \frac{E}{\hbar^2\omega^2}\Big[\sum_{n=-\infty}^\infty
\delta\Big(\frac{E}{\hbar\omega}-n\Big)-1\Big],
\end{aligned}
\end{equation}
In the second step, we have used the Poisson resummation formula. After 
adding the classical density of states (\ref{eqnClassicalDOS}) we recover for
positive energies the exact quantum mechanical result as was the case for the
incommensurate harmonic oscillator.

\section{Semiclassical theory for the commensurate anisotropic harmonic oscillator}
\subsection{From the anisotropic to the isotropic oscillator}
\label{sec:anisotoiso}
We now turn to the case of different but commensurate frequencies where
isolated periodic orbits coexist with a family of periodic orbits.
A key point for understanding the properties of a commensurate anisotropic
harmonic oscillator, both with respect to group theory and to its
semiclassical treatment, is the mapping to the isotropic harmonic
oscillator. With the spatial symmetries in mind, it might come as a
surprise that such a mapping should be possible because an anisotropic
harmonic oscillator does not obey the rotational symmetry characteristic
for the isotropic case. In fact, real space symmetries are not useful
here and we will be required to consider symmetries in phase space.

To motivate the relation between the commensurate anisotropic harmonic 
oscillator and the isotropic harmonic oscillator, we consider the energy 
spectrum of a harmonic oscillator with commensurate frequencies satisfying
$k_1\omega_1=k_2\omega_2=\omega$, where $k_1$ and 
$k_2$ are assumed to be integers without a common divisor. It is instructive 
to express the quantum numbers $\nu_i$ appearing in the 
energy spectrum (\ref{eqnQmEnergies}) in the commensurate case as 
$\nu_i = n_ik_i+\lambda_i$, where $\lambda_i=0,1,\dots,k_i-1$ and 
$n_i = 0, 1, 2,\dots$,\cite{louck73} so that the energy spectrum takes the form
\begin{equation}
E = \hbar\omega\Big(n_1 + n_2 + \frac{\lambda_1}{k_1} +
\frac{\lambda_2}{k_2} +\frac{k_1 + k_2}{2k_1k_2}\Big).
\label{eqnQmEnergiesc-AHO}
\end{equation}
These eigenenergies can be viewed as describing $k_1k_2$ sets of spectra of 
a two-dimensional isotropic harmonic oscillator shifted in energy relative 
to each other so that the isotropic harmonic oscillator appears to be 
represented multiple times in the commensurate anisotropic harmonic 
oscillator. This interpretation motivates us to seek a
mapping between these two systems and also to find their classical 
counterparts.

\begin{figure}[t]
\includegraphics[width=0.9\columnwidth]{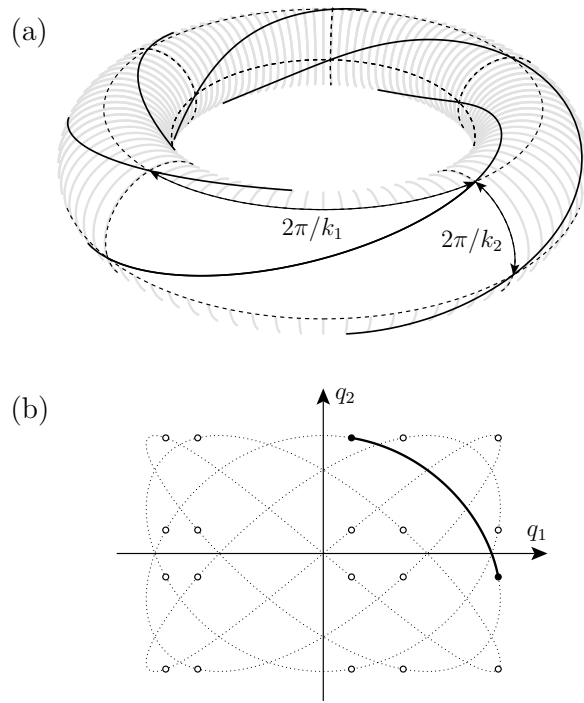}
\caption{\label{fig:torus}The motion of a commensurate anisotropic harmonic oscillator
with $k_1=5$ and $k_2=4$ is depicted in action-angle variables and in
spatial coordinates. (a) The full line represents a trajectory of the
commensurate anisotropic harmonic oscillator on a torus in phase space parametrized by
action-angle variables. The white section indicates one of the $k_1k_2$
elementary cells delineated by the dashed lines. Each cell after
identification of its opposite borders corresponds to an isotropic harmonic
oscillator. (b) The projection of the trajectory onto real space leads to a
Lissajous curve. The small circles indicate the positions where the
trajectory crosses from one elementary cell into another one. The full line
represents the trajectory in one such cell.}
\end{figure}

Because we are interested in the phase space structure, we briefly recall the 
results of the Hamilton-Jacobi method applied to the harmonic oscillator.\cite{golds80} The action $W_i$ is given by Eq.~(\ref{eqnAction}) and
the corresponding angle can be expressed as
\begin{equation}
\vartheta_i = \arctan\Big(\frac{m\omega q_i}{k_ip_i}\Big).
\label{eqnAngle}
\end{equation}
The motion of the harmonic oscillator thus takes place on a
torus that is parametrized by the angles $\vartheta_1$ and
$\vartheta_2$. The size of the torus is determined by the actions $W_1$
and $W_2$. The solid line in Fig.~\ref{fig:torus}(a) depicts the motion on
the torus for a commensurate anisotropic harmonic oscillator with
$k_1=5$ and $k_2=4$.

According to Eq.~(\ref{eqnAction}), the mapping from a commensurate
anisotropic
harmonic oscillator to an isotropic harmonic oscillator can be achieved by
the transformation\cite{louck73,moshi75,krame75}
\begin{subequations} \label{eqnCanonicalTransformation}
\begin{align} W_i &\mapsto \widetilde W_i = \frac{W_i}{k_i}\\
\vartheta_i & \mapsto \widetilde\vartheta_i = k_i \vartheta_i.
\end{align}
\end{subequations}
The definition of the new angles $\widetilde\vartheta_i$ ensures that the 
transformation is canonical. Remarkably, the different angles $\vartheta_i + 
\lambda_i 2 \pi/k_i$, $\lambda_i = 0,1,\ldots k_i-1$ are mapped onto the
same angle $\widetilde\vartheta_i$, because we have to identify 
$\widetilde\vartheta_i + 2\pi$ with $\widetilde\vartheta_i$ as usual. 
The transformation (\ref{eqnCanonicalTransformation}) thus leads to 
a folding of phase space into $k_i$ sheets for each mode $i$. This folding
is necessary to turn the commensurate anisotropic harmonic oscillator into an
isotropic harmonic oscillator, because it ensures that the original periods
$T_i = 2\pi k_i$ of the normal modes are mapped onto equal periods
$\widetilde T = 2\pi$. After the canonical transformation
(\ref{eqnCanonicalTransformation}) we now have
$k_1k_2$ isotropic harmonic oscillators instead of one commensurate
anisotropic harmonic oscillator.
Figure~\ref{fig:torus}(a) illustrates the transformation
(\ref{eqnCanonicalTransformation}). The white cell covering the interval 
$2\pi/k_i$ for angle $\vartheta_i$ turns into a square of side length $2\pi$
by the transformation (\ref{eqnCanonicalTransformation}).
By identifying opposite sides of the cell, a new torus is formed
on which the motion of an isotropic harmonic oscillator in phase space takes
place. Because the original torus contains $k_1k_2$ such cells marked by the
dashed lines, Eq.~(\ref{eqnCanonicalTransformation}) provides a mapping
from a commensurate anisotropic harmonic oscillator to 
$k_1k_2$ isotropic harmonic oscillators. 

\subsection{Symmetries of the commensurate anisotropic harmonic oscillator}
\label{sec:Symmetriesc-AHO}
We can now analyze the symmetries present in the commensurate anisotropic
harmonic oscillator: There exists a discrete symmetry associated with the 
mapping from the commensurate anisotropic harmonic oscillator to
multiple copies of isotropic harmonic oscillators. In addition, according
to Sec.~\ref{sectIHO}, each 
isotropic harmonic oscillator possesses the continuous symmetry SU(2).
The presence of the discrete symmetry is a direct consequence of the folding of
phase space discussed in Sec.~\ref{sec:anisotoiso}. We can introduce
complex variables $z_i = \sqrt{W_i} \exp(-i\vartheta_i)$
to write the transformation (\ref{eqnCanonicalTransformation}) as
\begin{equation}
\tilde z_i = \frac{|z_i|}{\sqrt{k_i}} \Big(\frac{z_i}{|z_i|}\Big)^{k_i},
\end{equation}
which is invariant under $z_i \mapsto z_i \exp(2\pi i/k_i)$. Hence, the
symmetry operations consist of rotations in the complex plane by multiples of
$2\pi/k_i$. The rotations form a symmetry group $\mathcal{C}_i = \{c_{{\rm
g},i},c^2_{{\rm g},i},\ldots,c^{k_i}_{{\rm g},i}\}$ generated by the fundamental
rotation $c_{{\rm g},i}$ by $2\pi/k_i$. The identity is given by the group element 
$c^{k_i}_{{\rm g},i}$ which leads to a rotation by $2\pi$. Taking together the 
rotations for both oscillator modes, we arrive at the product group 
$\mathcal{C} = \mathcal{C}_1\times\mathcal{C}_2$ containing $k_1 k_2$ elements and 
generated by the element $c_{\rm g} = c_{{\rm g},1} \times c_{{\rm g},2}$. For later 
use we note that the irreducible representations\cite{remark2,tinkh64} of 
$\mathcal{C}$ can be labeled by a tupel $(\lambda_1,\lambda_2)$ with $\lambda_i =
0,1,\ldots,k_i-1$. The character\cite{remark2} $\mathcal{X}$ of the group element 
$c_{\rm g}$ in the representation $(\lambda_1,\lambda_2)$ is given by
$\mathcal{X}_{\lambda_1,\lambda_2}(c_{\rm g}) = \exp (2\pi i \zeta_{\lambda_1,\lambda_2})$,
where we have introduced the phases
\begin{equation} \label{eqnPhaseInCharacterOfGroupElement}
\zeta_{\lambda_1,\lambda_2} = \frac{\lambda_1}{k_1} + \frac{\lambda_2}{k_2}.
\end{equation}

We emphasize that here we are concerned with a phase space symmetry. In
real space, a symmetry operation would map a segment of a Lissajous curve
such as the solid curve in Fig.~\ref{fig:torus}(b) onto one of the other
segments (delimited in the figure by two small circles) of
the same Lissajous curve. By starting from a fundamental segment, we can
arrive at the complete Lissajous curve by successively applying the
generator $c_{\rm g}$ of the group $k_1k_2$ times. In the Bloch sphere
picture, each segment is represented as a point on a Bloch sphere
corresponding to one isotropic harmonic oscillator. Because the
commensurate anisotropic harmonic oscillator has been mapped onto $k_1k_2$
isotropic harmonic oscillators, a complete Lissajous curve corresponds 
to one point on each of $k_1k_2$ Bloch spheres. The mapping between two
segments of a Lissajous curve then implies a jump from one Bloch sphere 
to another one.

\subsection{Construction of the semiclassical density of states}
For the semiclassical treatment of the commensurate anisotropic harmonic 
oscillator, we need to know which periodic orbits come in families and which 
ones are isolated. We start with a general Lissajous curve where both
one-dimensional harmonic oscillators along the $q_1$- and $q_2$-axes have
to return to the starting point of their oscillations at the same time.
Because their fundamental periods are equal to 
$T_i = 2\pi k_i/\omega$ and $k_1$ and $k_2$ are assumed to have no common 
divisor, the fundamental period for a Lissajous curve is given by 
$T = 2\pi k_1 k_2/\omega$. A Lissajous curve in configuration
space corresponds to a periodic orbit in full phase space where
$\mathsf{X}_1(2\pi k_1 k_2) = \mathsf{X}_2(2\pi k_1 k_2) = \openone$. By varying
the energy difference between the two oscillators and the relative phase of 
their oscillations, all possible Lissajous curves are obtained, including their 
degeneracies into oscillations along the coordinate axes. 

During the time $T$, $k_2$ oscillations along the $q_1$ axis and 
$k_1$ oscillations along the $q_2$ axis occur. Hence, all
oscillations along the $q_1$ ($q_2$) axis with fundamental period $T_1$
($T_2$) that are not repeated $k_2$ ($k_1$) times or multiples thereof are
not members of the family of Lissajous curves. They cannot be deformed
continuously into a Lissajous curve nor into each other and lie isolated in
phase space. Accordingly, for the isolated periodic orbits, one of the
matrizants $\textsf{X}_1$ or $\textsf{X}_2$ becomes equal to unity but not 
both.

These different types of periodic orbits lead to a decomposition of the
semiclassical density of states of the commensurate anisotropic harmonic 
oscillator into four terms 
\begin{equation} \label{eqnScDOSc-AHO}
\rho_{\rm sc}(E) = \bar \rho(E) + \delta \rho_\Gamma(E) +
\delta \rho_1(E) + \delta\rho_2(E),\quad (E>0)
\end{equation}
where $\bar \rho$ is the classical density of states
(\ref{eqnClassicalDOS}). $\delta \rho_\Gamma$ accounts for the contribution 
of the family of Lissajous curves denoted by $\Gamma$ here and in the
following, and the last two terms arise from the contribution of the
isolated periodic orbits. The latter can be treated in the same way as the
two periodic orbits of the incommensurate harmonic oscillator in 
Sec.~\ref{sec:i-AHO}. From Eq.~(\ref{eqnScDosi-AHO}) we find after
substituting the corresponding period and the Maslov index
\eqref{eqnMaslovIndexi-AHO}:
\begin{equation} \label{eqnScDosAHICFIsolatedLibration}
\delta \rho_1(E) = \frac{k_1}{\hbar \omega}\sum_{r=1}^{\infty}\strut'(-1)^r 
\frac{\sin\Big(\displaystyle k_1 r \frac{2\pi
E}{\hbar\omega}\Big)}{\sin(\pi rk_1/k_2)}.
\end{equation}
The prime indicates that the sum is restricted to the isolated
periodic orbits, that is, those repetitions $r$ for which the
denominator does not vanish; $\delta \rho_2(E)$ is obtained by
interchanging $k_1$ and $k_2$ in Eq.~(\ref{eqnScDosAHICFIsolatedLibration}).

To determine the contribution $\delta \rho_\Gamma$ of the family of 
Lissajous curves to the density of states, we make use of the discrete 
symmetry presented in Sec.~\ref{sec:Symmetriesc-AHO}. The mapping to the 
isotropic harmonic oscillator will allow us to take advantage of the results 
of Sec.~\ref{sectIHO}. 

The presence of a symmetry imposes a structure on the dynamical equations
that lets us consider the kinematics in a reduced space. In a quantum
mechanical picture, the full Hilbert space of the problem can be decomposed
into invariant subspaces corresponding to the irreducible representations of
the underlying symmetry group.\cite{tinkh64} As a consequence, the solution
of the Schr\"odinger equation that is restricted to these subspaces yields
symmetry-reduced spectra. 

A semiclassical treatment of this idea was used in
Refs.~\onlinecite{lauri91}--\onlinecite{robbi89} for discrete symmetries in 
configuration space and in Ref.~\onlinecite{creag93} for continuous 
symmetries. Although the commensurate anisotropic harmonic oscillator exhibits a discrete
as well as a continuous symmetry, we use symmetry reduction only for the
discrete group $\mathcal{C}$. After we have determined the symmetry reduced
densities of states, we will sum over all irreducible representations.
It can be generally shown that the full spectrum is constructed in
this way.\cite{robbi89,creag93,doll04}

\begin{figure}[t]
\includegraphics[width=\columnwidth]{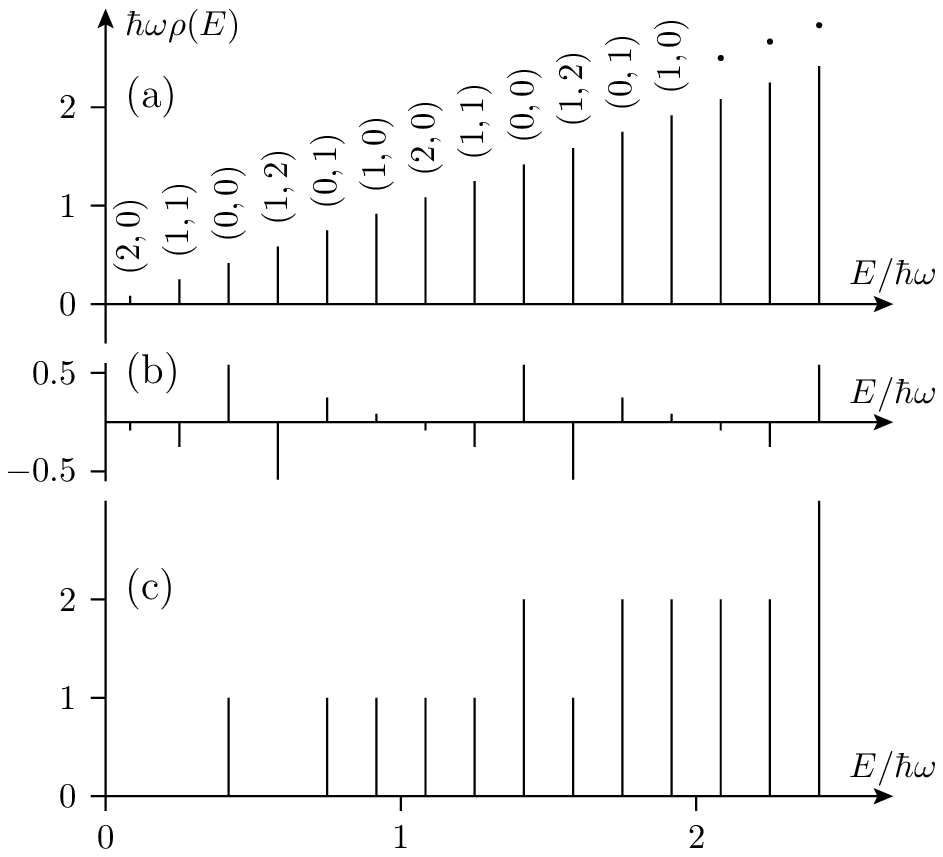}
\caption{\label{fig:dos}Density of states of the commensurate anisotropic harmonic
oscillator with frequency ratio 3:2. (a) Contribution of the classical
density of states and the family of periodic orbits. The six different
symmetry classes $(\lambda_1,\lambda_2)$ that repeat themselves
periodically are indicated. (b) The contribution arising from the isolated
periodic orbits. (c) The total density of states of the commensurate
anisotropic harmonic
oscillator.}
\end{figure}

An orthogonal projection corresponding to the irreducible representation
$(\lambda_1,\lambda_2)$ can be achieved using the projection operator
\cite{tinkh64} 
\begin{equation}
\label{eq:projector}
P_{(\lambda_1,\lambda_2)} = \frac{d_{\lambda_1,\lambda_2}}{|\mathcal{C}|}
\sum_{c} \mathcal{X}^\ast_{\lambda_1,\lambda_2}\!(c)\:U(c),
\end{equation} 
where $U(c)$ is the unitary operator describing the action of the group element 
$c$ in Hilbert space. $d_{\lambda_1,\lambda_2}$ is given by the dimension of 
the irreducible representation which in our case is always equal to one.\cite{remark2} By 
this projection the Green function $G(E)$, its trace and, in view of
Eq.~(\ref{eqnDiracIdentity}), the density of states are modified 
accordingly.\cite{doll04} The reduced densities of states corresponding to
the irreducible representation $(\lambda_1,\lambda_2)$ are found to be
\begin{align} \label{eqnSCDOSc-AHOReduced}
\delta \rho_\Gamma^{(\lambda_1,\lambda_2)}(E)&\\
&\hspace{-4em}= \frac{2 E}{\hbar^2\omega^2} \sum_{r=1}^\infty
\cos \Big[2\pi r \Big(\frac{E}{\hbar\omega} -
\zeta_{\lambda_1,\lambda_2} \Big)- (\tilde\sigma_r+1)
\frac{\pi}{2}\Big].\nonumber
\end{align}
The phases $\zeta_{\lambda_1,\lambda_2}$ are defined in
Eq.~(\ref{eqnPhaseInCharacterOfGroupElement}) and arise from the character
$\mathcal{X}$ present in Eq.~(\ref{eq:projector}). As we will see in the
discussion of Fig.~\ref{fig:dos}, this phase is responsible for a relative
shift of the reduced energy spectra.

The expression (\ref{eqnSCDOSc-AHOReduced}) can also be interpreted within
a classical picture, where the available phase space is reduced to a 
primitive cell, for example, the white area depicted in
Fig.~\ref{fig:torus}(a). The motion in full phase space may be recovered
using symmetry operations. Correspondingly, a Lissajous curve of the 
commensurate anisotropic harmonic oscillator is reduced to a fundamental 
segment, for example, the full line in Fig.~\ref{fig:torus}(b). The endpoints 
of such a segment are identified, thus yielding a periodic orbit of the 
isotropic harmonic oscillator whose continuous SU(2) symmetry still remains. 
We therefore can directly apply the results of Sec.~\ref{sectIHO}. It turns
out that the contribution to the semiclassical density of states of the
$r$th repetition of such an orbit acquires an additional phase given by the
character of the group element
$c_{\rm g}^r$. This group operation is needed to close the 
corresponding partial orbit in full phase space and, within a classical
picture, can be viewed as the origin of the shift of the reduced energy
spectra mentioned previously.

It remains to determine the Maslov index 
$\tilde\sigma_r$ of an orbit segment in Eq.~(\ref{eqnSCDOSc-AHOReduced}).
Let us consider as a representative of the whole family the oscillation
along the $q_1$ axis. We recall that only
$k_2$ repetitions of the fundamental periodic orbit form a member of
the family. We also have to keep in mind that we are now working in reduced 
dynamics where the $k_1 k_2$th part of a periodic orbit of a commensurate
anisotropic harmonic oscillator appears as a periodic orbit of an isotropic
harmonic oscillator. We therefore have to replace $r$ in
Eq.~(\ref{eqnMaslovIndexi-AHO}) by $r k_2/k_1 k_2$. If we add the correction 
$\Delta\sigma = -2$ as in the derivation of Eq.~(\ref{eqnMaslovIndexIHO}) for 
the isotropic harmonic oscillator, we obtain
\begin{equation}
\label{sigmarc-AHO}
\tilde\sigma_r = 2r\frac{k_1+k_2}{k_1k_2}-1.
\end{equation}

With Eqs.~(\ref{eqnClassicalDOS}),
(\ref{eqnScDosAHICFIsolatedLibration}), (\ref{eqnSCDOSc-AHOReduced}), 
and (\ref{sigmarc-AHO}) we have all the information needed to construct
the full density of states (\ref{eqnScDOSc-AHO}). We proceed in two steps as illustrated in Fig.~\ref{fig:dos} for the special case of $k_1=3$
and $k_2=2$. First, we note that the symmetry reduction performed for the family of
periodic orbits can also be applied to the classical density of states 
$\bar\rho(E)$. As we have seen, the reduced dynamics takes place in only
the $k_1k_2$th part of the original phase space volume. Correspondingly,
the classical density of states contributes to each reduced density of 
states only $\bar\rho(E)/k_1k_2$. We add this contribution to each of the
reduced densities of states arising from the family of periodic orbits and arrive at $k_1k_2$ series of delta peaks with a height increasing
linearly with energy. The density of states resulting from these two
contributions is shown in Fig.~\ref{fig:dos}(a) where $k_1k_2=6$ irreducible 
representations lead to a corresponding number of series of delta peaks  
shifted relative to each other in energy.

This density of states has to be corrected by the contribution 
$\delta \rho_1(E) + \delta \rho_2(E)$ arising from the isolated orbits 
(see Fig.~\ref{fig:dos}(b)). The sum of all contributions is 
shown in Fig.~\ref{fig:dos}(c), where we can check that the eigenenergies
with their degeneracies are correctly generated. We remark that the
first two peaks of Fig.~\ref{fig:dos}(a) are canceled by the first
two peaks in Fig.~\ref{fig:dos}(b), so that the ground state energy $5
\hbar\omega/12$ is associated with the irreducible representation (0,0).

Although this example indicates that the semiclassical treatment
reproduces the spectrum of the commensurate anisotropic harmonic oscillator, it is
useful to explicitly perform the sum over the reduced densities of
states (\ref{eqnSCDOSc-AHOReduced})
\begin{equation} \label{eqnSumOverAllIR}
\delta \rho_\Gamma(E) = \sum_{\lambda_1 = 0}^{k_1-1}
\sum_{\lambda_2=0}^{k_2-1}
\delta \rho_\Gamma^{(\lambda_1,\lambda_2)}(E).
\end{equation}
By interchanging the double sum with the sum over all repetitions $r$ in
Eq.~(\ref{eqnSCDOSc-AHOReduced}), it is seen that Eq.~(\ref{eqnSumOverAllIR})
vanishes except when $r$ is a multiple of $k_1 k_2$. Therefore only such orbits 
of the reduced dynamics actually contribute, which are repeated $k_1 k_2$ 
times. As
expected, these orbits are exactly the full Lissajous curves.

We may now replace the summation index $r$ in
Eq.~(\ref{eqnSCDOSc-AHOReduced}) by $r = k_1 k_2 k$ where $k = 1,2,\ldots$,
and arrive at
\begin{align}
\delta \rho_\Gamma(E)&\\ 
&\hspace{-1em}= \frac{2 E k_1 k_2}{\hbar^2\omega^2}
\sum_{k=1}^\infty (-1)^{k(k_1+k_2)} \cos\Big(kk_1k_2\frac{2\pi
E}{\hbar\omega}\Big). \nonumber
\end{align}
Adding the classical density of states and the contributions of the isolated
orbits, Eq.~(\ref{eqnScDOSc-AHO}) yields the exact density of states of the
commensurate anisotropic harmonic oscillator.\cite{brack95} This result can
be checked by summing the exact partition function and taking the inverse 
Laplace transform as we did for the incommensurate harmonic oscillator in 
Sec.~\ref{sec:i-AHO}.

\section{Conclusions}
The harmonic oscillator is a unique system with fundamental applications in
many areas in physics. It admits exact solutions that are often helpful in
the illustration of theoretical approaches and methods. As we have seen, it
is particularly interesting to consider a semiclassical treatment where a
relation between classical and quantum properties can be established. Beyond
the interesting different scenarios for periodic orbits
provided by the two-dimensional harmonic oscillator, this system
beautifully illustrates the relation between classical phase space
symmetries and quantum degeneracies.

\section{Suggested problems}
\textit{Problem 1}.
The conserved quantities $J_k$ of the isotropic harmonic oscillator defined
in Eq.~(\ref{eqnConservedQuantities}) can be interpreted as generators of
infinitesimal transformations. For example, $J_2$ generates rotations around 
the axis perpendicular to the $x_1$-$x_2$ plane. To better
understand the significance of $J_1$, consider its action on the energies 
$H_1$ and $H_2$ of the two modes, that is, evaluate the Poisson brackets 
$\{H_i,J_1\}$. How does $J_1$ modify an orbit? Hint: The ratio of
$H_1$ and $H_2$ is related to the eccentricity of the orbit, see 
Ref.~\onlinecite{mcint59} for a discussion.

It is also instructive to consider the $J_k$ as quantum-mechanical operators. 
Express them in terms of creation and annihilation operators $b_i^\dagger$
and $b_i$, respectively, for the two modes. Determine the action of $J_1\pm
iJ_2$ on an eigenstate $\vert \nu_1, \nu_2\rangle$ with $H_i\vert \nu_1,
\nu_2\rangle = \hbar\omega(\nu_i+1/2)\vert\nu_1,\nu_2\rangle$.

\textit{Problem 2}.
For the one-dimensional harmonic oscillator a transformation of the density
of states analogous to but simpler than the transition from Eq.~(\ref{eq:exactdos}) 
to Eqs.~(\ref{eqnClassicalDOSPlusFluctuations})--(\ref{eqnMaslovIndexi-AHO})
can be performed. From the partition function $\mathcal{Z}_\beta$ of the
one-dimensional harmonic oscillator, determine the density of states by an
inverse Laplace transformation
\begin{equation} \rho(E) = \frac{1}{2\pi
i}\!\int_{\epsilon-i\infty}^{\epsilon+i\infty}d \beta 
\mathcal{Z}_\beta \exp(\beta E),
\end{equation}
where $\epsilon> 0$. Use residue calculus and close the contour in the
left complex half plane. The resultant expression for the density of states
should contain the classical contribution $\bar\rho = 1/\hbar\omega$ and an oscillating part
\begin{equation}
\delta \rho(E) = \frac{2}{\hbar\omega} \sum_{r=1}^\infty \cos\Big(2\pi r
\Big(\frac{E}{\hbar\omega} - \frac{1}{2}\Big)\Big).
\end{equation}
How can this result be interpreted in terms of the classical properties of the
one-dimensional oscillator? Discuss the differences between the one- and
two-dimensional case.


\begin{thebibliography}{99}
\bibitem{remark}
The semiclassical approximation discussed here should be distinguished from
the semiclassical description where a quantum system is coupled to an
electromagnetic or gravitational field that is treated classically.
\bibitem{remark2}
A (matrix) representation of a symmetry group is a mapping $D$ from the group
elements $c$ to square matrices $D(c)$. The matrices lead to the same multiplication
table as the corresponding group elements. The character
$\mathcal{X}$ of a group element in a certain representation is defined as the
trace of the corresponding matrix. The groups $\mathcal{C}_i$ discussed here
are cyclic and therefore abelian. As for all abelian groups\cite{tinkh64} there are
$|\mathcal{C}_i| = k_i$ irreducible representations of $\mathcal{C}_i$. These
are one-dimensional so that the matrices degenerate to numbers and thus coincide 
with their character. Because of
$D^{k_i}(c_{{\rm g},i}) = D(c^{k_i}_{{\rm g},i}) = 1$, we obtain for $c_{{\rm g},
i}$ the irreducible representations $D_{\lambda_i}(c_{{\rm g},i}) = 
\exp\{2\pi i \lambda_i/k_i\}$, labeled by $\lambda_i = 0,1,\ldots,k_i-1$.
\end{thebibliography}
\end{document}